...

# Invisibility of antisymmetric tensor fields in the light of $F(R)$ gravity

Ashmita Das[*]

*Department of Physics, Indian Institute of Technology, North Guwahati, Guwahati, Assam 781039, India*

Tanmoy Paul[†] and Soumitra SenGupta[‡]

*Department of Theoretical Physics, Indian Association for the Cultivation of Science,
2A & 2B Raja S.C. Mullick Road, Kolkata 700 032, India*

A natural question arises from observable signatures of scalar, fermion, and vector degrees of freedom (d.o.f.) in our Universe along with spin 2 symmetric tensor field in the form of gravity: why is our Universe is free of any perceptible signature of massless antisymmetric tensor modes? This work brings out a natural explanation of these phenomena through higher curvature quantum d.o.f. in the gravity sector that were dominant in the early universe. In the backdrop of a $F(R)$ gravity model, we propose how the scalar d.o.f. associated with higher curvature term in the model can generate a heavily suppressed coupling between any antisymmetric massless modes and various standard model fields.

A surprising feature in the present Universe is that there is no noticeable observable effect of antisymmetric tensor fields in any natural phenomena. Apart from being a massless representation of the underlying Lorentz group, such higher rank antisymmetric tensor fields also appear as a massless closed string mode in a heterotic string model [1] and have considerable interests in the context of String theory. In this context, the second rank antisymmetric tensor field, known as the Kalb-Ramond (KR) field [2], has been studied extensively. From dimensional consideration it is easy to argue that the coupling of KR field to matter should be $\sim 1/M_p$ (where $M_p$ is the fundamental scale of the gravity, namely the Planck scale), which is the same as the coupling of gravity with matter. But there has been no experimental evidence of the footprint of the KR field on the present Universe. All experimental efforts within their precision in the domain of cosmological and astrophysical experiments so far have produced only negative results in terms of detecting any signatures of antisymmetric tensor fields. This indicates that if such tensor fields exist, it must be severely suppressed at the present scale of the Universe. Thus the question that arises is as follows: why are the effects of the Kalb-Ramond field or any higher rank antisymmetric tensor field less perceptible than the force of gravitation? Attempts have been made to answer the above question in the room of extradimensional braneworld model where the effect of such antisymmetric tensor fields are diluted on the visible brane through the exponential warping of the spacetime geometry [3–6].

However, in the present work, we aim to show that without bringing in any extra dimension, the origin of such suppression of the KR as well as the other higher rank antisymmetric fields can be explained in the light of higher curvature $F(R)$ theory in four spacetime dimensions. It is well known that Einstein-Hilbert action can be generalized by adding higher order curvature terms, which naturally arise from diffeomorphism invariance of the action. Such terms appear in string theory from quantum corrections. $F(R)$ [7–18], Gauss-Bonnet [19–21], or more generally Lanczos-Lovelock gravity [22–24] are some of the candidates in higher curvature gravitational theory. While the Gauss-Bonnet and Lanczos-Lovelock gravitational theories have nontrivial consequences only in higher dimension (higher than 4), $F(R)$ gravity has a significant role even in four spacetime dimensions. However only for some specific choices of $F(R)$ (for which $F'(R) > 0$), it can be made as free of ghost.

The field strength tensor of any massless rank $n$ antisymmetric tensor field $X_{a_1 a_2 \ldots a_n}$ can be expressed as

$$Y_{a_1 a_2 \ldots a_{n+1}} = \partial_{[a_{n+1}} X_{a_1 a_2 \ldots a_n]}.$$

In four-dimensional spacetime, the rank of antisymmetric tensor field can at most be 3, beyond which the corresponding field strength tensor vanishes identically.

[*]ashmita@iitg.ernet.in
[†]pul.tnmy9@gmail.com
[‡]tpssg@iacs.res.in

We begin our discussion with rank 2 antisymmetric KR field. The action of massless KR field along with spin $\frac{1}{2}$ fermion and gauge field in a background $F(R)$ gravity in four dimensions can be written as

$$S = \int d^4x\sqrt{-g}\left[\frac{F(R)}{2\kappa^2} - \frac{1}{4}H_{\mu\nu\rho}H^{\mu\nu\rho} + \bar{\Psi}i\gamma^\mu D_\mu \Psi\right.$$
$$-\frac{1}{4}F_{\mu\nu}F^{\mu\nu} - \frac{1}{M_p}\bar{\Psi}\gamma^\mu\sigma^{\nu\rho}H_{\mu\nu\rho}\Psi$$
$$\left.-\frac{1}{M_p}A^{[\mu}F^{\nu\rho]}H_{\mu\nu\rho}\right], \tag{1}$$

where $H_{\mu\nu\rho}$ ($= \partial_{[\mu}B_{\nu\rho]}$) is the field strength tensor of the KR field $B_{\mu\nu}$ and $\frac{1}{2\kappa^2} = M_p^2$. The third and fourth terms of the above action are the kinetic Lagrangians of spin $\frac{1}{2}$ fermion fields (symbolized by $\Psi$) and $U(1)$ gauge field (symbolized by $A_\mu$), respectively, with $\gamma^\mu$ representing the gamma matrix satisfying $\{\gamma^\mu, \gamma^\nu\} = 2g^{\mu\nu}$ and $D_\mu$ ($= \partial_\mu + ieA_\mu$) being the covariant derivative. Further the last two terms denote the coupling of $B_{\mu\nu}$ with the fermion and the gauge field, respectively. These interaction terms play the key role in finding some observable signatures of the KR field.

Introducing an auxiliary field $A(x)$, the above action (1) can be equivalently written as

$$S = \int d^4x\sqrt{-g}\left[\frac{1}{2\kappa^2}(F'(A)(R - A) + F(A))\right.$$
$$-\frac{1}{4}H_{\mu\nu\rho}H^{\mu\nu\rho} + \bar{\Psi}i\gamma^\mu D_\mu\Psi - \frac{1}{4}F_{\mu\nu}F^{\mu\nu}$$
$$\left.-\frac{1}{M_p}\bar{\Psi}\gamma^\mu\sigma^{\nu\rho}H_{\mu\nu\rho}\Psi - \frac{1}{M_p}A^{[\mu}F^{\nu\rho]}H_{\mu\nu\rho}\right]. \tag{2}$$

By the variation of the auxiliary field $A(x)$, one easily obtains $A = R$. Plugging back this solution $A = R$ into action (2), initial action (1) can be reproduced. At this stage, perform a conformal transformation of the metric as

$$g_{\mu\nu}(x) \to \tilde{g}_{\mu\nu}(x) = e^{-\sqrt{\frac{2}{3}}\kappa\xi(x)}g_{\mu\nu}(x)$$

$\mu, \nu$ run form 0 to 3. $\xi(x)$ is the conformal factor and related to the auxiliary field as $e^{-\sqrt{\frac{2}{3}}\kappa\xi} = F'(A)$. Because of such conformal transformation, the gamma matrices $\gamma^\mu$ and the spin connection $\sigma^{\nu\rho}$ transform as

$$\gamma^\mu \to \tilde{\gamma}^\mu = e^{\frac{1}{2}\sqrt{\frac{2}{3}}\kappa\xi}\gamma^\mu$$

and

$$\sigma^{\nu\rho} \to \tilde{\sigma}^{\nu\rho} = e^{\sqrt{\frac{2}{3}}\kappa\xi}\sigma^{\nu\rho},$$

respectively. Using the above expressions along with the aforementioned relation between $\xi(x)$ and $A(x)$, one ends up with the following scalar-tensor action:

$$S = \int d^4x\sqrt{-\tilde{g}}\left[\frac{\tilde{R}}{2\kappa^2} + \frac{1}{2}\tilde{g}^{\mu\nu}\partial_\mu\xi\partial_\nu\xi - \left(\frac{AF'(A) - F(A)}{2\kappa^2 F'(A)^2}\right)\right.$$
$$-\frac{1}{4}e^{-\sqrt{\frac{2}{3}}\kappa\xi}H_{\mu\nu\rho}H_{\alpha\beta\delta}\tilde{g}^{\mu\alpha}\tilde{g}^{\nu\beta}\tilde{g}^{\rho\delta} - \frac{1}{4}F_{\mu\nu}F_{\alpha\beta}\tilde{g}^{\mu\alpha}\tilde{g}^{\nu\beta}$$
$$+ e^{\sqrt{\frac{2}{3}}\kappa\xi}\Psi^+\tilde{\gamma}^0 i\tilde{\gamma}^\mu D_\mu \Psi - \frac{1}{M_p}\Psi^+\tilde{\gamma}^0\tilde{\gamma}^\mu\tilde{\sigma}^{\nu\rho}H_{\mu\nu\rho}\Psi$$
$$\left.-\frac{1}{M_p}e^{-\sqrt{\frac{2}{3}}\kappa\xi}A_{[\alpha}F_{\beta\delta]}H_{\mu\nu\rho}\tilde{g}^{\mu\alpha}\tilde{g}^{\nu\beta}\tilde{g}^{\rho\delta}\right], \tag{3}$$

where $\tilde{R}$ is the Ricci scalar formed by $\tilde{g}_{\mu\nu}$. The field $\xi(x)$ is a scalar field with a potential $\frac{AF'(A)-F(A)}{2\kappa^2 F'(A)^2}$ [$= V(A(\xi))$, say]. Thus the higher curvature degree of freedom (d.o.f.) manifests itself through the scalar field d.o.f. $\xi(x)$ with a potential $V(\xi)$ that in turn depends on the form of $F(R)$. Further it is evident that due to the appearance of the scalar field $\xi(x)$ (from higher curvature d.o.f.), the kinetic terms of the fermion field and the KR field become noncanonical while the electromagnetic field remains still canonical. In order to make such kinetic terms canonical, we redefine the fields as follows:

$$B_{\mu\nu} \to \tilde{B}_{\mu\nu} = e^{-\frac{1}{2}\sqrt{\frac{2}{3}}\kappa\xi}B_{\mu\nu}$$
$$\Psi \to \tilde{\Psi} = e^{\frac{1}{2}\sqrt{\frac{2}{3}}\kappa\xi}\Psi$$

and

$$A_\mu \to \tilde{A}_\mu = A_\mu.$$

However, in terms of redefined fields, the above action (3) can be expressed in Einstein frame as

$$S = \int d^4x\sqrt{-\tilde{g}}\left[\frac{\tilde{R}}{2\kappa^2} + \frac{1}{2}\tilde{g}^{\mu\nu}\partial_\mu\xi\partial_\nu\xi - V(\xi)\right.$$
$$-\frac{1}{4}\tilde{H}_{\mu\nu\rho}\tilde{H}^{\mu\nu\rho} - \frac{1}{4}\tilde{F}_{\mu\nu}\tilde{F}^{\mu\nu} + \bar{\tilde{\Psi}}i\tilde{\gamma}^\mu\tilde{D}_\mu\tilde{\Psi}$$
$$-\frac{1}{M_p}e^{(-\frac{1}{2}\sqrt{\frac{2}{3}}\kappa\xi)}\bar{\tilde{\Psi}}\tilde{\gamma}^\mu\tilde{\sigma}^{\nu\rho}\tilde{H}_{\mu\nu\rho}\tilde{\Psi}$$
$$-\frac{1}{M_p}e^{(-\frac{1}{2}\sqrt{\frac{2}{3}}\kappa\xi)}\tilde{A}_{[\alpha}\tilde{F}_{\beta\delta]}\tilde{H}_{\mu\nu\rho}$$
$$\left.+ \text{terms proportional to}\partial_\mu\xi\right], \tag{4}$$

where $\tilde{D}_\mu = \partial_\mu + ie\tilde{A}_\mu$. It may be observed that the interaction terms (between $\tilde{B}_{\mu\nu}$ and $\tilde{\Psi}$, $\tilde{A}_\mu$) of the canonical scalar-tensor action [see Eq. (4)] carry an exponential factor $e^{(-\frac{1}{2}\sqrt{\frac{2}{3}}\kappa\xi)}$ over the usual gravity-matter coupling $1/M_p$. As mentioned earlier, the scalar field potential $V(\xi)$ depends on the form of $F(R)$. However in general, the stability of $V(\xi)$ follows from the following two conditions on $F(R)$:

$$[2F(R) - RF'(R)]_{<R>} = 0$$

and

$$\left[\frac{F'(R)}{F''(R)} - R\right]_{<R>} > 0.$$

In order to achieve an explicit expression of a stable scalar potential, we first consider the form of $F(R)$ as an exponential analytic function of the Ricci scalar,

$$F(R) = R + \alpha(e^{-\beta R} - 1), \quad (5)$$

where $\alpha$ and $\beta$ are the free parameters (with mass dimensions $[\beta] = -2$ and $[\alpha] = +2$) of the theory. This model is considered as one of the strong and viable candidates in the field of $F(R)$ gravity for reasons such as the following: (1) this model of $F(R)$ gravity is free of ghosts for $\alpha\beta < 1$, (2) it satisfies $\lim_{R\to 0}[F(R) - R] = 0$, which indicates that there exists a flat spacetime solution, and (3) such an exponential model has important consequences in the context of inflationary cosmology as well as late time acceleration as discussed in [16,17].

However, as we illustrate later, some other generic forms of $F(R)$ models also lead to the similar conclusions as obtained for the model described in Eq. (5).

For the specific choice of $F(R)$ shown in Eq. (5), the potential $V(\xi)$ becomes

$$V(\xi) = \frac{\alpha}{2\kappa^2} e^{(2\sqrt{\frac{2}{3}}\kappa\xi)}$$
$$\times \left[\frac{(1 - e^{-\sqrt{\frac{2}{3}}\kappa\xi})}{\alpha\beta}\left(\ln\left[\frac{1 - e^{-\sqrt{\frac{2}{3}}\kappa\xi}}{\alpha\beta}\right] + 1\right) + 1\right]. \quad (6)$$

This potential has a minimum at

$$\langle\xi\rangle = \frac{1}{\kappa}\sqrt{\frac{3}{2}}\ln\left[\frac{1}{1 - \alpha\beta}\right]. \quad (7)$$

Equation (7) indicates that for a wide range of values of the product $\alpha\beta$ (between 0 and 1), the vacuum expectation value (vev) of the scalar field $\xi(x)$ becomes of the order of $\frac{1}{\kappa}$ ($\sim 10^{19}$ GeV). This suggests that at the early epoch of the Universe, when the energy scale was high, the scalar field $\xi$ was a dynamical scalar d.o.f. giving rise to new interaction vertices with fermion and gauge fields. However as the Universe evolved into a lower energy scale due to cosmological expansion, $\xi$ finally froze into its vacuum expectation value $\langle\xi\rangle$ as given in Eq. (7). The action in Eq. (4) therefore turns out to be

$$S = \int d^4x\sqrt{-\tilde{g}}\left[\frac{\tilde{R}}{2\kappa^2} - \frac{1}{4}\tilde{H}_{\mu\nu\rho}\tilde{H}^{\mu\nu\rho} - \frac{1}{4}\tilde{F}_{\mu\nu}\tilde{F}^{\mu\nu}\right.$$
$$+ \bar{\tilde{\Psi}}i\tilde{\gamma}^\mu\tilde{D}_\mu\tilde{\Psi} - \frac{1}{M_p}e^{(-\frac{1}{2}\sqrt{\frac{2}{3}}\kappa\langle\xi\rangle)}\bar{\tilde{\Psi}}\tilde{\gamma}^\mu\tilde{\sigma}^{\nu\rho}\tilde{H}_{\mu\nu\rho}\tilde{\Psi}$$
$$\left. - \frac{1}{M_p}e^{(-\frac{1}{2}\sqrt{\frac{2}{3}}\kappa\langle\xi\rangle)}\tilde{A}_{[\alpha}\tilde{F}_{\beta\delta]}\tilde{H}^{\mu\nu\rho}\right], \quad (8)$$

where the terms proportional to $\partial_\mu\xi$ vanish as $\xi(x)$ is frozen at its vev. The last two terms in the above expression of action give the coupling of KR field to fermion, $U(1)$ gauge field and are given by

$$\lambda_{\text{KR-fermion}} = \frac{1}{M_p}e^{(-\frac{1}{2}\sqrt{\frac{2}{3}}\kappa\langle\xi\rangle)}$$
$$= \frac{1}{M_p}\sqrt{1 - \alpha\beta} \quad (9)$$

and

$$\lambda_{\text{KR-}U(1)} = \frac{1}{M_p}e^{(-\frac{1}{2}\sqrt{\frac{2}{3}}\kappa\langle\xi\rangle)}$$
$$= \frac{1}{M_p}\sqrt{1 - \alpha\beta}, \quad (10)$$

respectively. The above two equations clearly demonstrate that the product $\alpha\beta$ must be less than unity; otherwise the couplings of the KR field become imaginary, an unphysical situation. However the condition $\alpha\beta < 1$ is also supported by the fact that the higher curvature terms may have their origin from quantum corrections, which from dimensional argument are suppressed by Planck scale.

Equations (9) and (10) indicate that the coupling strengths of KR field to matter fields are heavily suppressed [as $\langle\xi\rangle$ is positive, see Eq. (7)] over the usual gravity-matter coupling strength $1/M_p$. Such suppression was also reported in the context of the Randall-Sundrum higher dimensional model [3]. Here, on the other hand the suppression originates from higher curvature d.o.f. even in four dimensions irrespective of choosing the background geometry. This may explain why the present Universe is dominated by spacetime curvature and carries practically no observable signature of the rank 2 antisymmetric Kalb-Ramond field (or equivalently the torsion field).

Let us now consider the rank 3 antisymmetric tensor field $X_{\alpha\beta\rho}$ with the corresponding field strength tensor $Y_{\alpha\beta\rho\delta}$ ($= \partial_{[\alpha}X_{\beta\rho\delta]}$). The action for such a field in four dimensions is

$$S[X] = \int d^4x\sqrt{-g}Y_{\alpha\beta\rho\delta}Y^{\alpha\beta\rho\delta}.$$

Adopting the same procedure as for the KR field, one can end up with the coupling of the field $X$ to matter in the canonical scalar-tensor action as

$$\Omega_{X\text{-fermion}} = \frac{1}{M_p} e^{(-\frac{1}{2}\sqrt{\frac{2}{3}}\kappa\langle\xi\rangle)}$$
$$= \frac{1}{M_p}\sqrt{1-\alpha\beta} \quad (11)$$

and

$$\Omega_{X\text{-}U(1)} = \frac{1}{M_p} e^{(-\sqrt{\frac{2}{3}}\kappa\langle\xi\rangle)}$$
$$= \frac{1}{M_p}(1-\alpha\beta), \quad (12)$$

where $\Omega_{X\text{-fermion}}$ and $\Omega_{X\text{-}U(1)}$ denote the coupling between the X-fermion and X-$U(1)$ gauge field, respectively. Like the case of KR field, $\Omega_{X\text{-fermion}}$ and $\Omega_{X\text{-}U(1)}$ are also suppressed in comparison to $1/M_p$. However $\lambda_{\text{KR-fermion}}$ and $\Omega_{X\text{-fermion}}$ carry the same suppression factor while the interaction with the electromagnetic field becomes progressively smaller with the increasing rank of the tensor field. Therefore the visibility of an antisymmetric tensor field in our present Universe becomes lesser with the increasing rank of the tensor field.

TABLE I. $\langle\xi\rangle$ and the couplings of rank 2 (onwards) antisymmetric field to matter fields for a wide range of $\alpha\beta$.

| $(1-\alpha\beta)$ | $\langle\xi\rangle$ | $\lambda_{\text{KR-fermion}}$ | $\lambda_{\text{KR-}U(1)}$ | $\Omega_{X\text{-fermion}}$ | $\Omega_{X\text{-}U(1)}$ |
|---|---|---|---|---|---|
| $10^{-2}$ | $\sim M_p$ | $\sim 10^{-1}/M_p$ | $\sim 10^{-1}/M_p$ | $\sim 10^{-1}/M_p$ | $\sim 10^{-2}/M_p$ |
| $10^{-4}$ | $\sim M_p$ | $\sim 10^{-2}/M_p$ | $\sim 10^{-2}/M_p$ | $\sim 10^{-2}/M_p$ | $\sim 10^{-4}/M_p$ |
| $10^{-8}$ | $\sim 10 M_p$ | $\sim 10^{-4}/M_p$ | $\sim 10^{-4}/M_p$ | $\sim 10^{-4}/M_p$ | $\sim 10^{-8}/M_p$ |

It may be observed that $\langle\xi\rangle$ [see Eq. (7)] and correspondingly the couplings depend on the product of the parameters $\alpha$ and $\beta$. Using the explicit expression of $\langle\xi\rangle$ obtained in Eq. (7), Table I clearly reveals that all the couplings are suppressed by an additional exponential factor over the gravitational coupling $1/M_p$. The suppression increases as the value of $\alpha\beta$ tends to unity.

At this stage, it deserves mentioning that beside the model presented in Eq. (5), there exist several other $F(R)$ models for which the intrinsic scalar d.o.f. suppresses the coupling strengths of antisymmetric tensor fields with matter fields by an additional reduced factor over the gravity-matter coupling $1/M_p$. In fact any generic form of $F(R)$ that corresponds to a stable scalar potential with a vev of Planck order leads to the suppression of antisymmetric tensor fields in the present context. Some such $F(R)$ models are (1) $F(R) = R - \gamma\ln(R/\mu^2) - \delta R^2$ [25] where $\gamma$, $\mu$, and $\delta$ are constants, (2) $F(R) = a[e^{bR}-1]$ where $a$, $b$ are constants, and (3) $F(R) = R + \omega R^2 + \rho[e^{-\sigma R}-1]$ ($\omega$, $\rho$, $\sigma$ are constants). With appropriate choices of the parameters, these $F(R)$ models also lead to the suppression on the coupling strengths between antisymmetric tensor fields and various matter fields.

In conclusion, one important feature in $F(R)$ gravity is the intrinsic existence of an extra scalar d.o.f., besides the massless graviton. Such a scalar field appears with a potential that depends on the form of $F(R)$. Here, we consider the form of $F(R)$ as an exponential function of Ricci scalar in four dimensions for which the scalar field acquires a stable value. It turns out that this vacuum expectation value of the scalar field suppresses the coupling of all antisymmetric tensor fields to the matter fields over the gravity-matter coupling strength $1/M_p$. The suppression actually increases with the rank of the tensor field. In this context it may be noted that the Lagrangian of a higher spin field represented by a higher rank symmetric tensor violates gauge invariance under gauge transformation of the tensor field when it couples to a curved background [26]. Since the suppression of antisymmetric tensors has been shown to originate from the higher curvature terms in an arbitrary curved background, a similar argument for suppression of higher spin fields cannot be extended here due to the undesirable loss of gauge invariance. It has already been demonstrated that such a loss of gauge invariance does not happen for the antisymmetric tensor fields coupled to curved background. Therefore, this may well serve as an explanation of why the large scale behavior of our present Universe is solely governed by gravity and carries practically no observable footprints of antisymmetric tensor fields.

## ACKNOWLEDGMENTS

SSG acknowledges the financial support from SERB Research Project No. EMR/2017/001372.

[1] M. J. Duff and J. T. Liu, Phys. Lett. B **508**, 381 (2001); O. Lebedev, Phys. Rev. D **65**, 124008 (2002).
[2] M. Kalb and P. Ramond, Phys. Rev. D **9**, 2273 (1974).
[3] B. Mukhopadhyaya, S. Sen, and S. SenGupta, Phys. Rev. Lett. **89**, 121101 (2002).
[4] B. Mukhopadhyaya, S. Sen, and S. SenGupta, Phys. Rev. D **76**, 121501(R) (2007).

[5] A. Das, B. Mukhopadhyaya, and S. SenGupta, Phys. Rev. D 90, 107901 (2014).

[6] A. Das and S. SenGupta, Phys. Lett. B 698, 311 (2011).

[7] T. P. Sotiriou and V. Faraoni, Rev. Mod. Phys. 82, 451 (2010).

[8] A. De Felice and S. Tsujikawa, Living Rev. Relativity 13, 3 (2010).

[9] A. Paliathanasis, Classical Quantum Gravity 33, 075012 (2016).

[10] A. Das, H. Mukherjee, T. Paul, and S. SenGupta, Eur. Phys. J. C 78, 108 (2018).

[11] N. Banerjee and T. Paul, Eur. Phys. J. C 77, 672 (2017).

[12] S. Bahamonde, S. D. Odintsov, V. K. Oikonomou, and M. Wright, Ann. Phys. (Amsterdam) 373, 96 (2016).

[13] S. SenGupta and S. Chakraborty, Eur. Phys. J. C 76, 552 (2016).

[14] S. Anand, D. Choudhury, Anjan A. Sen, and S. SenGupta, Phys. Rev. D 92, 026008 (2015).

[15] G. Cognola, E. Elizalde, S. Nojiri, S. D. Odintsov, L. Sebastiani, and S. Zerbini, Phys. Rev. D 77, 046009 (2008).

[16] S. D. Odintsov, D. Saez-Chillon Gomez, and G. S. Sharov, Eur. Phys. J. C 77, 862 (2017).

[17] E. V. Linder, Phys. Rev. D 80, 123528 (2009).

[18] K. Bamba, C. Q. Zeng, and C. C. Lee, J. Cosmol. Astropart. Phys. 08 (2010) 021.

[19] E. Elizalde, S. D. Odintsov, V. K. Oikonomou, and T. Paul, arXiv:1810.07711.

[20] S. Nojiri, S. D. Odintsov, and O. G. Gorbunova, J. Phys. A 39, 6627 (2006).

[21] G. Cognola, E. Elizalde, S. Nojiri, S. D. Odintsov, and S. Zerbini, Phys. Rev. D 73, 084007 (2006).

[22] S. Chakraborty, T. Paul, and S. SenGupta, arXiv:1804.03004.

[23] C. Lanczos, Z. Phys. 73, 147 (1932); Ann. Math. 39, 842 (1938).

[24] D. Lovelock, J. Math. Phys. (N.Y.) 12, 498 (1971).

[25] S. Nojiri and S. D. Odintsov, Phys. Rep. 505, 59 (2011).

[26] N. Bouatta, G. Compere, and A. Sagnotti, arXiv:hep-th/0409068; R. Rahman and M. Taronna, arXiv:1512.07932; C. Fornsdal, Phys. Rev. D 18, 3624 (1978); M. Vasiliev, arXiv:hep-th/9910096v2; V. E. Didenko and E. D. Skvortsov, arXiv:1401.2975v5.